\documentclass[]{pasj01}
\draft

\begin{document} 
\Received{}
\Accepted{}

\title{Hierarchical Tree Algorithm for Collisional N-body Simulations on GRAPE}

\author{Toshiyuki \textsc{Fukushige}\altaffilmark{1}%
}
\altaffiltext{1}{K\&F Computing Research Co.}
\email{fukushig@kfcr.jp}

\author{Atsushi \textsc{Kawai},\altaffilmark{1}}
\email{kawai@kfcr.jp}


\KeyWords{galaxies: star clusters --- methods: n-body simulations --- stellar dynamics} 

\maketitle

\begin{abstract}


We present an implementation of the hierarchical tree algorithm on the
individual timestep algorithm (the Hermite scheme) for collisional
$N$-body simulations, running on GRAPE-9 system, a special-purpose
hardware accelerator for gravitational many-body simulations.  Such
combination of the tree algorithm and the individual timestep
algorithm was not easy on the previous GRAPE system mainly because its
memory addressing scheme was limited only to sequential access to a
full set of particle data. The present GRAPE-9 system has an indirect
memory addressing unit and a particle memory large enough to store all
particles data and also tree nodes data. The indirect memory
addressing unit stores interaction lists for the tree algorithm, which
is constructed on host computer, and, according to the interaction
lists, force pipelines calculate only the interactions necessary.  In
our implementation, the interaction calculations are significantly
reduced compared to direct $N^2$ summation in the original Hermite
scheme.  For example, we can archive about a factor 30 of speedup
(equivalent to about 17 teraflops) against the Hermite scheme for a
simulation of $N=10^6$ system, using hardware of a peak speed of 0.6
teraflops for the Hermite scheme.

\end{abstract}

\section{Introduction}



Collisional $N$-body simulations, in which the equations of motion of
$N$ particles integrated numerically, have been extensively used in
studies of dense star clusters, such as globular cluster, open
cluster, and clusters with black holes, and also in studies of
planetary formation. One feature of the collisional $N$-body
simulations is need for relatively high accuracy in the force
calculations, because the total number of timestep is very large to
simulate relatively long simulation span, such as the relaxation
timescale.  Another feature is a wide difference in orbital timescale
of particles since two particles can approach arbitrary close.  The
individual timestep algorithm, first developed by Aarseth
(\cite{a63}), has been a powerful tool that handles the collisional
$N$-body system, whose basic idea is to assign different times and
timesteps to particles in the system.

GRAPE(GRAvity PipE)(\cite{s+91}) is a special purpose hardware that
can accelerate the individual timestep algorithm.  GRAPE hardware has
specialized pipelines for the gravitational-force calculation, which
is the most expensive part of the collisional $N$-body
simulations. Among the individual timestep algorithm, the Hermite
scheme (\cite{ma92}) can efficiently use the GRAPE hardware, in which
the block individual timestep algorithm (\cite{m86}) and the 4th-other
Hermite integration are used. GRAPE-6 (\cite{m+03}) is a
massive-parallel hardware for the collisional $N$-body simulations using
the Hermite scheme. It consists of 1728 pipeline chips and has a peak
speed of around 64 teraflops.

Although direct summation algorithm was used for the force
calculations on the GRAPE-6 system, whether it is the really best
solution or not remains a question. The Barnes-Hut tree algorithm
(\cite{bh86}) is one of algorithms that reduces the calculation cost
by replacing forces from distant particles by those from a virtual
particle at their center of mass. McMillan and Aarseth (\cite{ma93})
have demonstrated that it is possible to implement a combination of
the Barnes-Hut tree algorithm and the individual timestep algorithm
that runs efficiently on a single-processor computers. However, on the
GRAPE-6 system, the combination of the tree algorithm and the
individual timestep algorithm was not possible, because its memory
addressing scheme was limited only to sequential access to a full set
of particle data, and there is not enough memory size for particle
data.

We successfully implemented the combination of the tree algorithm and
the individual timestep algorithm on GRAPE-9 system. GRAPE-9 is a
newly-developed system that uses FPGA(Field Programmable Gate Array)
device and the force and predictor pipelines the same as GRAPE-6 chip
are integrated in the device. The GRAPE-9 system also has an indirect
memory addressing unit and a relatively large-sized particle memory,
implemented by widely-used DRAM device.  Interaction lists for the
tree algorithm can be stored in the GRAPE-9 system, and the force
pipelines can calculate only the interactions necessary. By our
implementation, the interaction calculations are significantly reduced
from the direct summation in the Hermite scheme.

The plan of this paper is as follows. In section 2 we describe
implementation of the tree algorithm on the Hermite scheme using
GRAPE-9. In section 3, we present the performance and accuracy of our
implementation. Section 4 is for discussion.

\section{Implementation}

In this section, we describe how the interactions are calculated using
the tree algorithm in our implementation.  In an ideal way, the
interaction list should be created at every block timestep using
predicted particle data, but it is not practical. In our
implementation, the tree structure and the interaction lists are
created only at intervals of $\Delta t_{\rm tree}$, and the same
interaction lists are used during $\Delta t_{\rm tree}$.  Therefore, the
interval $\Delta t_{\rm tree}$ becomes a cause of error in the
interactions calculation since the tree structure is deformed as time
advances. The interval $\Delta t_{\rm tree}$ has to be small enough not to
affect simulation results and its performance. In our implementation,
the maximum size of timestep is set to be $\Delta t_{\rm tree}$ for
simplicity.

With the original Hermite scheme, the previous GRAPE system (GRAPE-6
system) performs the integration of one step in the following way:

\begin{itemize}

\item[1.] As the initialization procedure, the host computer sends all data of 
all particles to the memory on GRAPE. 

\item[2.] The host computer selects particles to be integrated at the
  present system time.

\item[3.] Repeat 4-6 for all particles selected.

\item[4.] The host computer predicts the position and velocity of the particle, 
and sends them to GRAPE. 

\item[5.] GRAPE calculates the force from all other particles, and then returns the 
results to the host computer.

\item[6.] The host computer integrates the orbits of the particles and determines the 
new timestep. The updated particle data are sent to the memory on GRAPE.

\item[7.] The host computer updates the present system time and go back to step 2.

\end{itemize}

In our new implementation with the tree algorithm, the GRAPE system
(GRAPE-9 system) performs the integration of one step in the following
way (the bold item number shows the step that changes from the
original algorithm):

\begin{itemize}

\item[{\bf 1.}] At intervals of $\Delta t_{\rm tree}$ (and at
  initial), the host computer makes tree data. The procedure includes
  construction of a tree structure, identification of groups of
  particles for which the same interaction list is used by traversing
  the tree structure, and creation of interaction lists for the
  groups. The host computer sends all particles data, the
  interaction lists for all groups, and tree node data listed up in
  the interaction lists, to the memory on GRAPE. The tree node data
  are stored in the memory as (pseudo-)particles that have positions,
  velocities, accelerations, and their time derivatives.

\item[2.] The host computer selects particles to be integrated at the
  present system time.

\item[3.] Repeat 4-6 for all particles selected.

\item[4.] The host computer predicts the position and velocity of the particle, 
and sends them to GRAPE. 

\item[{\bf 4a.}] The host computer sends the index number of the
  interaction list for the particle to GRAPE.

\item[{\bf 5.}] GRAPE calculates the force from particles in the
  interaction list, and then returns the results to the host computer.

\item[6.] The host computer integrates the orbits of the particles and
  determines the new timestep. The updated particle data are sent to
  the memory on GRAPE.

\item[7.] The host computer updates the present system time and
  go back to step 2.

\end{itemize}

The differences from the original algorithm are in three steps: in
step 1, at intervals of $\Delta t_{\rm tree}$, the tree structure and
the interaction list are created and sent to GRAPE. In step 4a, the
index number of the interaction list is sent to GRAPE. In step 5, the
force are calculated from particles in the interaction list, instead
of from all particles.

In order to use efficiently the GRAPE hardware, we use the modified
tree algorithm (as already described in step 1), which was developed
by Barnes (\cite{b90}) and implemented on the GRAPE hardware by Makino
(\cite{m91}). With this algorithm, tree traversal is performed for a
group of neighboring particles and an interaction list is created for
the group. The maximum number of particles in the group, $n_{\rm
  crit}$, is set to be optimal at which the total computing time is
the minimum.  As we increase $n_{\rm crit}$, the interaction
calculation on GRAPE increases since interactions between particles in
a group are calculated directly and the interaction list becomes
longer.  On the other hand, as we decrease $n_{\rm crit}$, an
efficiency of usage of the GRAPE hardware becomes lower, since the
number of particles in the same group at each block step becomes
smaller on average. For the present system, $n_{\rm crit}=2000-4000$,
is close to optimal. Note that, with such $n_{\rm crit}$, interactions
with a rather large number of the neighboring particles, about $10^4$
(for $\theta = 0.5$), are directly calculated.  The part of the tree
algorithm in our code is almost the same as that used in the previous
studies (\cite{fkm05},\cite{yf05}). The simulation program is written
using the GRAPE-6 compatible API library and two additional functions
for step 1 and steps 4a-5, respectively.

\begin{figure}
 \begin{center}
\includegraphics[width=12cm]{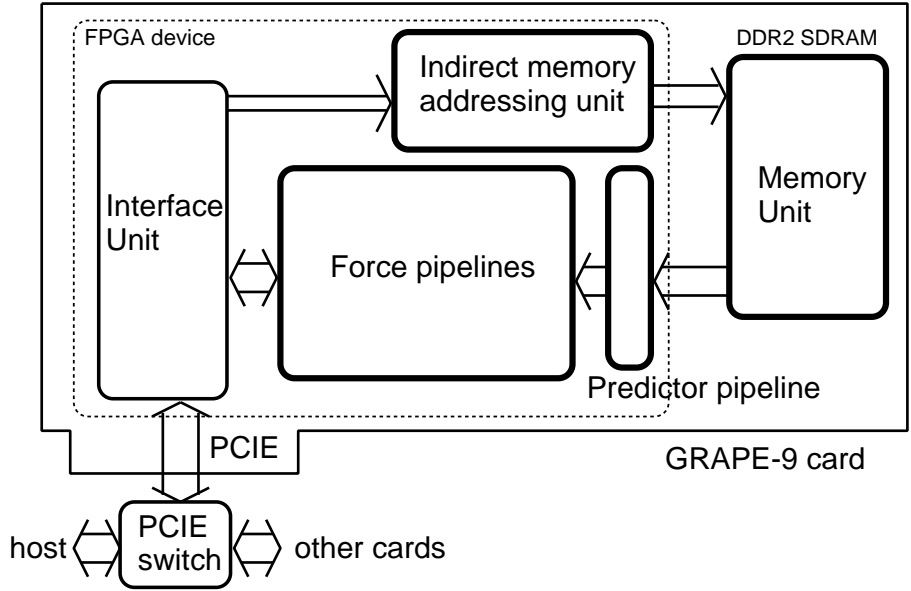} 
 \end{center}
\caption{Overall structure of the GRAPE-9 system}\label{fig:g9}
\end{figure}

We implemented these algorithm on the GRAPE-9 (model 5000) system. The
GRAPE-9 system consists of 8-16 GRAPE-9 cards, connected to the host
computer via a PCI Express switch device (PLX PEX8696). The GRAPE-9
card is a PCI Express extension card on which one FPGA(Field
Programmable Gate) device and one DDR2 SDRAM (SO-DIMM module) memory
are mounted.  In the FPGA device, the force and predictor pipelines,
almost identical to the GRAPE-6 chip, and an indirect memory
addressing unit are integrated, which is illustrated in Figure
\ref{fig:g9}.  The interaction lists for the tree algorithm is stored
in the indirect memory addressing unit of the FPGA device, actually in
on-chip memory of the FPGA device, and all particles data and tree
node data are stored in the memory unit, which consists of the DDR2
SDRAM memory.  According to the interaction list, the indirect memory
addressing unit outputs an address entry for the memory unit, and the
force pipelines calculate only the interactions necessary. We use
Altera Cyclone V 5CGXC9 for the FPGA device.  The wealthy amount of
the on-chip memory in this device is one of reasons that enables our
implementation.  For the present implementation, we use a
configuration in which GRAPE-6-compatible 14 force pipelines and one
predictor pipeline are integrated, which operates at 98MHz.  Other
details on the GRAPE-9 system will be discussed elsewhere.

\begin{figure}
 \begin{center}
 \includegraphics[width=14cm]{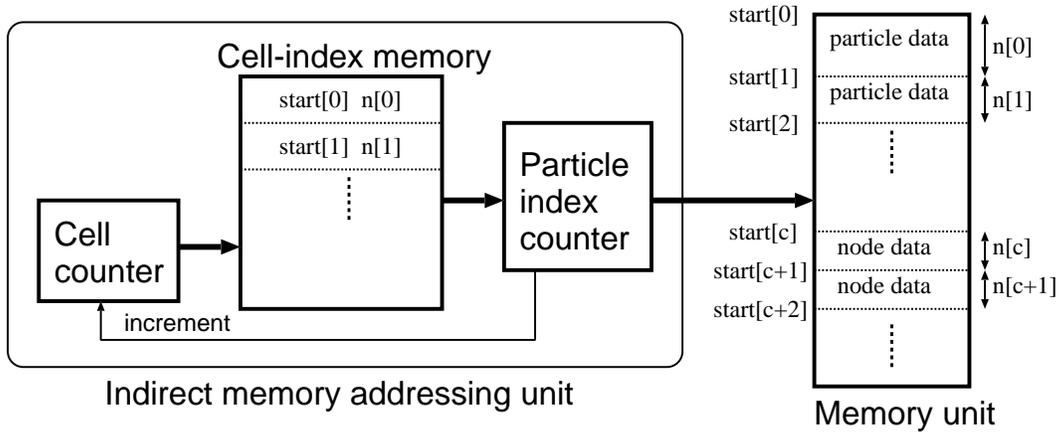} 
 \end{center}
\caption{Block diagram of the indirect memory addressing unit}\label{fig:indirect}
\end{figure}

As for the indirect memory addressing unit, we use the particle index
unit same as in GRAPE-5 (\cite{k+00}), which was designed for the
cell-index method (\cite{qb75}) to handle short-range forces in a
periodic boundary condition. Figure \ref{fig:indirect} shows a block
diagram for the indirect memory addressing unit. It consists of the
cell-index memory and two counters: the cell counter and the particle
index counter.  In the cell-index memory, sets of start address and
count number are stored. According to the output of the cell-index
memory, the particle index counter generates entries to the memory
unit.  The cell counter indicates address entry for the cell-index
memory.  Actually, in step 4a, the host computer sends the start
address and count number of the cell counter for the group of the
interaction list.  Because size of the on-chip memory of the FPGA
device is limited (about several Mbits), we store the interaction list
for the tree algorithm in such form, instead of full sets of indices.
The entry size of the cell-index memory is 98304 for the present FPGA
device. In order to reduce total length of the interaction lists in
the cell-index memory, we rearrange all particles in the Peano-Hilbert
order and store the tree node data for each group in a consecutive
location in the memory unit.  Since we typically use 1GB DDR2 SDRAM
SO-DIMM (8GB at maximum), the memory unit can store 10 millions
particles at the maximum for each card.

When we perform calculations using multiple GRAPE-9 cards, we use two
parallelization methods in combination: (1) multiple cards calculate
the force on the same set of ($i$-)particles, but from different set
of ($j$-)particles. (called $j$-parallel) (2) multiple cards calculate
the force on different set of ($i$-)particles whose interaction list
(group) are different. (called $i$-parallel)

\section{Performance and Accuracy}

In this section, we discuss performance and accuracy for our
implementation.  As the benchmark runs, we integrated the Plummer
model with equal-mass particles.  We use the standard unit
(\cite{hm86}) in which $M=G=-4E=1$. Here $M$ and $E$ are the total
mass and energy of the system, and $G$ is the gravitational
constant. The timestep criterion is that of Aarseth (\cite{a99}) with
$\eta=0.01$. For the softening parameter, we used an $N$-dependent
softening, $\varepsilon=1/N$.  We set $n_{\rm crit} = 4000$ for all
runs except for runs of $\theta=0.3$, $N$=512k and 1M ($n_{\rm
 crit} = 7000$ and $14000$), and the interval $\Delta t_{\rm tree} =
1/64$.

We used the GRAPE-9 system that consists of 8 GRAPE-9 cards and whose
peak speed is 630 Gflops. Here we count operations for the
gravitational force and its time derivatives as 57 floating-point
operations. Host computer has Intel core i7-3820 (4core, 3.6GHz)
CPU. Communications between the host computer and each GRAPE-9 card is
PCI Express gen1 4lane (1GB/s peak for each direction).  In order to
use 8 cards simultaneously, we used a parallelization method whose
degree is 4 for $j$-parallel and 2 for $i$-parallel.

\begin{figure}
 \begin{center}
  \includegraphics[width=9cm,angle=270]{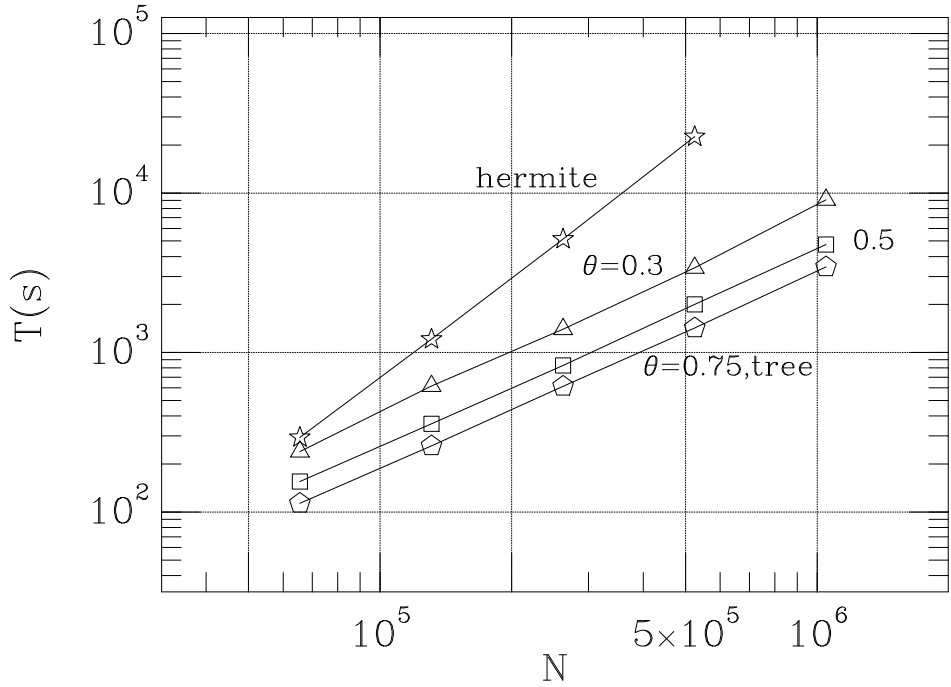} 
 \end{center}
\caption{
Calculation time, $T$, to integrate the system for one time unit as a
function of the number of particles, $N$. The triangle, square, and
pentagon indicate the results with opening angles for the tree
algorithm $\theta = 0.3, 0.5$ and $0.75$, respectively.  The star
indicates the result for the Hermite scheme.
}\label{fig:nt}
\end{figure}

Figure \ref{fig:nt} shows the calculation time, $T$, to integrate the
system for one time unit as a function of the number of particles,
$N$, for $\theta = 0.3, 0.5$ and $0.75$, where $\theta$ is a opening
parameter for the tree algorithm. For comparison, we also plot the
calculation time for the original Hermite scheme.  We measured the
calculation time, $T$, from simulation time $t=0.25$ to $0.5$
(and multiplied it by four) to avoid the complication due to the
startup procedure.

\begin{figure}
 \begin{center}
  \includegraphics[width=9cm,angle=270]{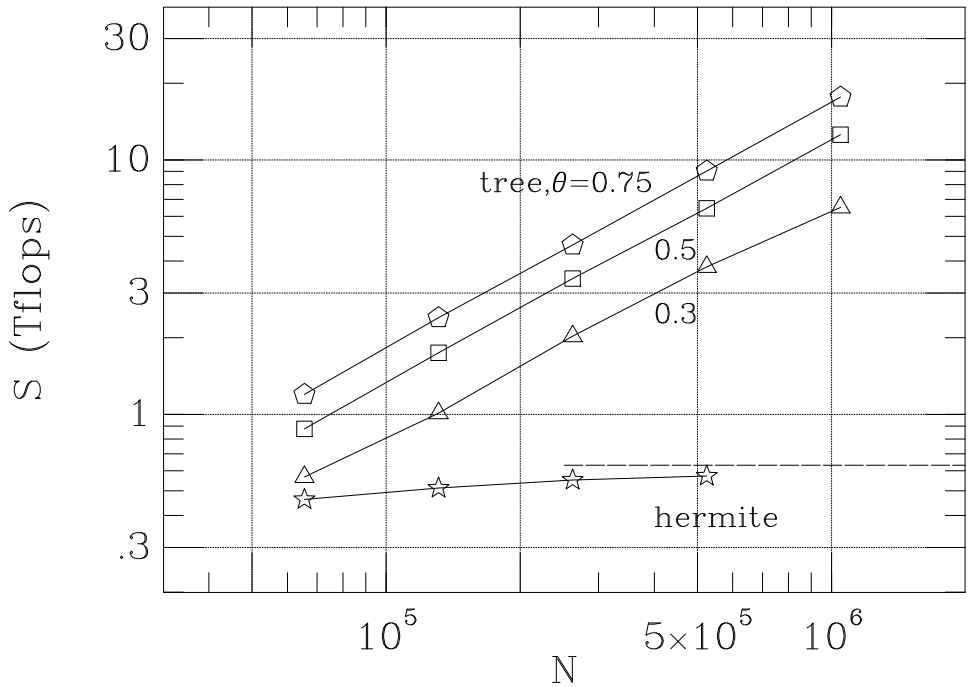} 
 \end{center}
\caption{ 
Equivalent-performance, $S$, defined in the text, as a function of
the number of particles, $N$. The triangle, square, and pentagon
indicate those with opening angles for the tree algorithm $\theta =
0.3, 0.5$ and $0.75$, respectively.  The star indicates that for the
Hermite scheme.  The thin dashed line indicates the peak performance
for the Hermite scheme.
}\label{fig:nsp}
\end{figure}

Figure \ref{fig:nsp} shows another plot with an equivalent-performance, $S$, 
defined by
\begin{equation}
S = {57Nn_{\rm step} \over T}{n_{\rm step,h} \over n_{\rm step}} 
= {57Nn_{\rm step,h} \over T},
\end{equation}
where $n_{\rm step}$ is the total number of individual timestep to
integrate one time unit, and $n_{\rm step,h}$ is that for the Hermite
scheme.  The equivalent-performance means the performance in the case
that we perform the same simulation within the same time using the
Hermite scheme.  For the ratio, $n_{\rm step,h}/n_{\rm step}$, of
$N=$1M, we used instead those of $N=$512k for each $\theta$. The ratio
$n_{\rm step,h}/n_{\rm step}$ itself is close to unity, for example,
$n_{\rm step,h}/n_{\rm step}= 1.042$ for $N=$512k, $\theta=0.75$,
which means even if we use the tree algorithm total number of
individual timestep does not increase so much. We can see that about a
factor ($N/30$k) of speedup (for $\theta = 0.5$) against the original
Hermite scheme is achieved.

\begin{figure}
 \begin{center}
  \includegraphics[width=9cm,angle=270]{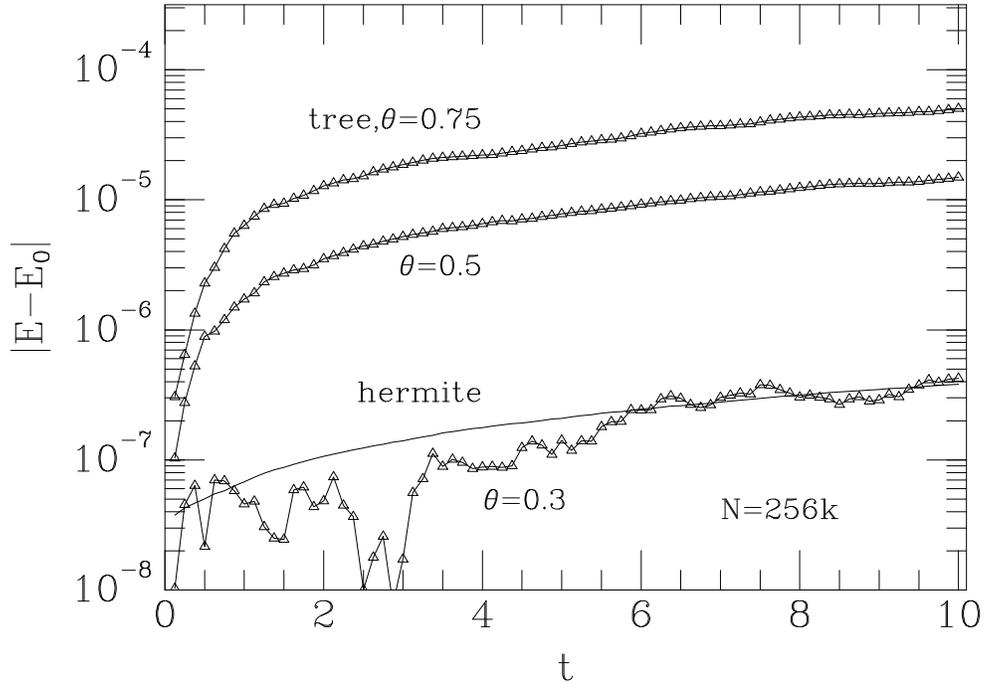} 
 \end{center}
\caption{
Time evolution of errors in total energy for the $N=256$k run.  The
triangle indicates those for the tree algorithm $\theta =0.3, 0.5$ and
$0.75$, and the sold curve indicates that for the Hermite scheme.
}\label{fig:err256k}
\end{figure}

\begin{figure}
 \begin{center}
  \includegraphics[width=9cm,angle=270]{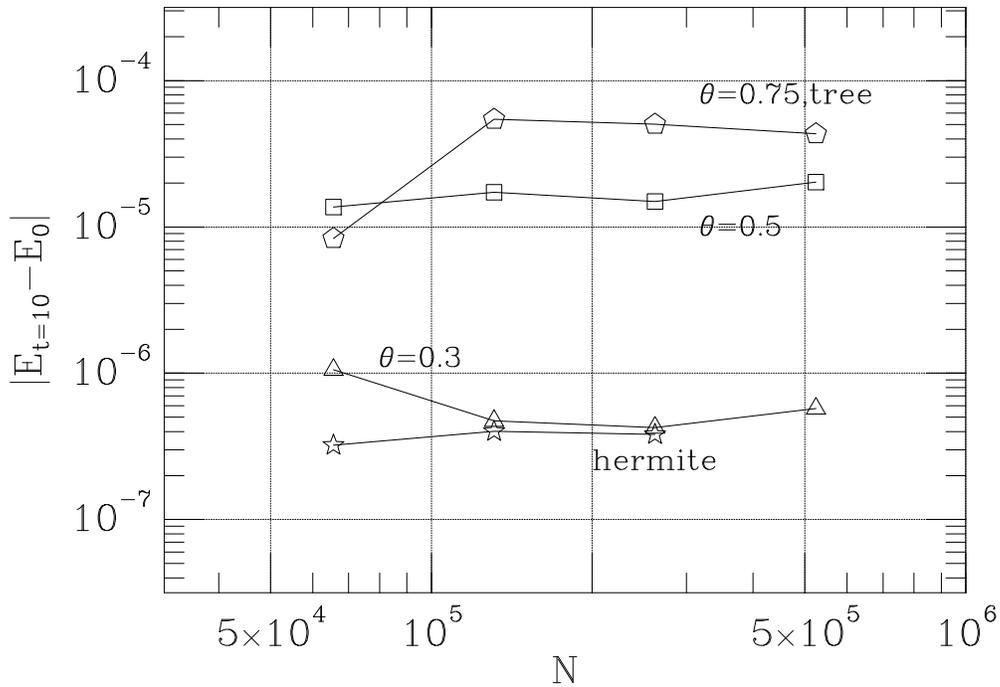} 
 \end{center}
\caption{
Errors in the total energy at simulation time $t=10$ as a function of
the number of particles, $N$.  The triangle, square, and pentagon
indicate those with opening angles for the tree algorithm $\theta =
0.3, 0.5$ and $0.75$, respectively.  The star indicates that for the
Hermite scheme.
}\label{fig:err}
\end{figure}

Figure \ref{fig:err256k} shows errors in the total energy as a function
of time up to simulation time $t=10$ for $N=256$k. The calculation of
the potential energy is obtained with the direct summation on GRAPE,
not with the tree algorithm. We can see that the errors in our
implementation of the tree algorithm increase linearly as time
advances. In figure \ref{fig:err}, the errors in the total energy at
time $t=10$ are summarized. From both figures, we can see that the errors
for $\theta = 0.3$ are comparable to those for the original Hermite
scheme.

\begin{table}
  \tbl{Breakdown of calculation time}{%
  \begin{tabular}{rr|rccccc}
  \hline
   $N$ & $\theta$ & ${\bar N_{\rm int}}$ & ${\bar n_{\rm i}}$ & $T_{\rm grape}$(s) & $T_{\rm tree}$(s) & $T_{\rm comm}$(s) & $T_{\rm host}$(s) \\ 
  \hline
   65536  & 0.75 & 5740  & 29.3 & $8.5\times 10^{-6}$ &$1.5\times 10^{-7}$ &$9.8\times 10^{-7}$ &$1.6\times 10^{-7}$ \\
   65536  & 0.5  & 11081 & 28.8 & $1.6\times 10^{-5}$ &$2.4\times 10^{-7}$ &$9.8\times 10^{-7}$ &$1.6\times 10^{-7}$ \\
   65536  & 0.3  & 22351 & 28.6 & $3.3\times 10^{-5}$ &$4.0\times 10^{-7}$ &$9.8\times 10^{-7}$ &$1.5\times 10^{-7}$ \\
   262144 & 0.75 & 6433  & 29.8 & $9.2\times 10^{-6}$ &$9.7\times 10^{-8}$ &$1.2\times 10^{-6}$ &$2.7\times 10^{-7}$ \\
   262144 & 0.5  & 12644 & 29.5 & $1.8\times 10^{-5}$ &$1.5\times 10^{-7}$ &$1.2\times 10^{-6}$ &$2.7\times 10^{-7}$ \\
   262144 & 0.3  & 28168 & 29.4 & $4.0\times 10^{-5}$ &$2.7\times 10^{-7}$ &$1.2\times 10^{-6}$ &$2.6\times 10^{-7}$ \\
   \hline
   \end{tabular}}\label{tab:timing}
\end{table}

We discuss breakdown of the calculations using a simple performance
model. The calculation time per one particle step is expressed as
\begin{equation}
T = T_{\rm grape}
+ T_{\rm tree} 
+ T_{\rm comm} 
+ T_{\rm host}. 
\end{equation}
In Table \ref{tab:timing}, the terms of the right-hand side measured
in actual runs are listed. The first term of the right-hand side,
$t_{\rm grape}$, is the time to calculate the force and its time
derivative for one particle on GRAPE-9, expressed as
\begin{equation}
T_{\rm grape} \simeq  
{\bar N_{\rm int}} t_{\rm pipe} \left({{\bar n_{\rm i}}\over n_{\rm pipe}}\right)^{-1}
\end{equation}
where ${\bar N_{\rm int}}$ is average numbers of the interaction list
for the tree algorithm. In the case of one GRAPE-9 card, $t_{\rm pipe}
= 7.6\times 10^{-10}$ (s). The factor ${\bar n_{\rm i}} / n_{\rm
  pipe}$ expresses a decrease in performance when the number of
particles that calculate interactions simultaneously (at step 5) is
less than $n_{\rm pipe}$.  Here, $n_{\rm pipe}$ and ${\bar n_{\rm i}}$
are the maximum and average number of the particles that calculate
interactions simultaneously, respectively. The number ${\bar n_i}$
becomes much smaller for the tree algorithm than in the original
Hermite scheme, because, even in the same block step, the particles
belong to several different groups.  The number ${\bar n_i} \sim 30$
in actual runs, which does not depends on $N$ nor $\theta$ so much. For
the present system, $n_{\rm pipe}=56$, since it has 14 real force
pipelines and each real pipeline serves as 4 virtual multiple
pipelines (\cite{m+97}).

The second term, $T_{\rm tree}$, is the time for the tree data
processing spent in the host computer (step 1), which are listed in
Table \ref{tab:timing} for $\Delta t_{\rm tree} = 1/64$.  The time
$T_{\rm tree}$ is proportional to $1/\Delta t_{\rm tree}$ and does not
depend on $N$ so much. The third term, $T_{\rm comm}$, expresses the
time to transfer data between the host computer and GRAPE, which
include data conversion.  Since about 200 byte data transfer are
required per one particle step, the sustained transfer speed is about
200MB/s.  The fourth term, $T_{\rm host}$, is the time for the host
computer to perform computations to integrate one particle other than
$T_{\rm tree}$.

As for the breakdown, at first, we note that, in the first term
$T_{\rm grape}$, the decrease in performance due to small ${\bar
  n_{\rm i}}$ is rather large and the sustained performance decreases
to about half of its peak performance. This is partly because $n_{\rm
  pipe}=56$ for the present system is not small enough.  The system
which has $n_{\rm pipe}$ less than 30 is desirable for our
implementation. Second, the largest term is $T_{\rm comm}$ among
three terms other than $T_{\rm grape}$, and the fraction to $T_{\rm grape}$
is not small compared in the original Hermite scheme because the
number of interaction ${\bar N_{\rm int}}$ is not large, of
course. Third, the time $T_{\rm tree}$ is small enough compared to
other terms, in the case of $\Delta t_{\rm tree} = 1/64$.

\section{Discussion}

We successfully implemented the hierarchical tree algorithm on the
individual timestep algorithm (the Hermite scheme) for collisional
$N$-body simulations on the GRAPE-9 system.  The present GRAPE-9
system has the indirect memory addressing unit and the memory unit
large enough to store all particles data and also tree nodes data.  In
our implementation, the interaction calculations are significantly
reduced, compared to direct $N^2$ summation in the original Hermite
scheme.

In comparison to other methods that also reduce calculation amounts
for the individual timestep algorithm successfully, our implementation
has one advantageous feature that interactions from particles at an
intermediate range are evaluated in more accurate way.  The neighbor
scheme (\cite{ac73}, \cite{na12}) is an example of such methods. In the
scheme, the force on a particle is divided into two components, the
neighbor force and the regular force and calculations amount are
reduced by evaluating the regular force less frequently. P$^3$T
(Particle-Particle Particle-Tree, \cite{ofm11}, \cite{ipm15}) is
another example. In P$^3$T, the force on a particle is split into
short-range and long-range contribution. The short-range force are
evaluated with the Hermite scheme and the long-range force are
evaluated with the tree algorithm and leapfrog integrator.  It is
reported that less accurate evaluation for intermediate range force
might influence the angular momentum evolution (see \cite{ipm15}).

At present, the GRAPE-9 system is probably good solution for the
implementation of the hierarchical tree algorithm on the individual
timestep algorithm. Further improvement with the next generation FPGA
device would provide more powerful computing systems.  Shipment within
one year of a new FPGA device (Altera Arria 10) that has more than 4
times number of logic elements, 3 times operation speed, 8 times data
transfer speed (PCIE gen3 8lane), 4 times size of the on-chip memory
and 10 times memory bandwidth (DDR3/DDR4 SDRAM), compare to the
current FPGA device (Altera Cyclone V), has been announced. New system
using such FPGA device would be able to provide about 10 times of
performance with keeping smaller $n_{\rm pipe}(\sim 30)$, which is
another required ingredient for our implementation.

Porting of our implementation on other accelerators, such as GPGPU
device, is presumably feasible and in preparation. Typically, very
large number of parallel operations must be executed on such
accelerator.  Since, for our implementation, the number of the
interaction calculations that can be executed in parallel becomes
smaller, some ingenuities would be necessary for an efficient use of
the accelerator.

\begin{ack}
We are grateful to Hiroshi Daisaka and Ataru Tanikawa for 
helpful discussions and variable comments on this study.
\end{ack}


\newpage

\begin{thebibliography}{}

\bibitem[Aarseth(1963)]{a63} Aarseth,~S.~J. 1963, MNRAS, 126, 223
\bibitem[Aarseth(1999)]{a99} Aarseth,~S.~J. 1999, Celest. Mech. Dyn. Astron., 73, 127
\bibitem[Ahmad \& Cohen(1973)]{ac73} Ahmad,~A. \& Cohen,~L. 1973, J. Comput. Phys, 12, 389
\bibitem[Barnes(1990)]{b90} Barnes,~J.~E. 1990, J. Comput. Phys, 87, 161
\bibitem[Barnes \& Hut(1986)]{bh86} Barnes,~J.~E., \& Hut,~P. 1986, Nature, 326, 446
\bibitem[Fukushige et al.(2005)]{fkm05} Fukushige,~T., Makino,~J., \& Kawai,~A. 2005, PASJ, 57, 1009
\bibitem[Heggie \& Mathieu(1986)]{hm86} Heggie,~D.~C., \& Mathieu,~R.~D. 1986, in The Use of Supercomputer in Stellar Dynamics, ed. P.Hut \& S.McMillan (New York: Springer), 233
\bibitem[Iwasawa et al.(2015)]{ipm15} Iwasawa,~M., Portegies Zwart,~S., \& Makino,~J. 2015, Computional Astrophysics and Cosmology, 2, 6
\bibitem[Kawai et al.(2000)]{k+00} Kawai,~A., Fukushige,~T., Makino,~J., \& Taiji,~M. 2000, PASJ, 52, 659
\bibitem[Makino(1991)]{m91} Makino,~J. 1991a, PASJ, 43, 621
\bibitem[Makino \& Aarseth(1992)]{ma92} Makino,~J. \& Aarseth,~S.~J. 1992, PASJ, 44, 141
\bibitem[Makino et al.(2003)]{m+03} Makino,~J., Fukushige,~T., Koga,~M., \& Namura,~K. 2003, PASJ, 55, 1163
\bibitem[Makino et al.(1997)]{m+97} Makino,~J., Taiji,~M., Ebisuzaki,~T., \& Sugimoto,~D. 1997, ApJ, 480, 432
\bibitem[McMillan(1986)]{m86} McMillan,~S.~L.~W. 1986, in The Use of Supercomputer in Stellar Dynamics, ed. P.Hut \& S.McMillan (New York: Springer), 156
\bibitem[McMillan \& Aarseth(1993)]{ma93} McMillan,~S.~L.~W. \& Aarseth,~S.~J. 1993, ApJ, 414, 200
\bibitem[Nitadori \& Aarseth(2012)]{na12} Nitadori,~K. \& Aarseth,~S.~J. 2012, MNRAS, 424, 545
\bibitem[Oshino et al.(2011)]{ofm11} Oshino,~S., Makino,~J., \& Funato,~Y. 2011, PASJ, 63, 881
\bibitem[Quentrec \& Brot(1975)]{qb75} Quentrec,~B. \& Brot,~C. 1975, J. Comput. Phys, 13, 430
\bibitem[Sugimoto et al.(1991)]{s+91} Sugimoto,~D., Chikada,~Y., Makino,~J., Ito,~T., Ebisuzaki,~T., \& Umemura,~M. 1990, Nature, 345, 33
\bibitem[Yoshikawa \& Fukushige(2005)]{yf05} Yoshikawa,~K. \& Fukushige,~T. 2005, PASJ, 57, 849



\end{thebibliography}
\end{document}